\documentclass[english]{iopart}
\usepackage[T1]{fontenc}
\usepackage{color}
\usepackage{babel}

\usepackage[unicode=true,pdfusetitle,
 bookmarks=true,bookmarksnumbered=false,bookmarksopen=false,
 breaklinks=true,pdfborder={0 0 1},backref=false,colorlinks=true]
 {hyperref}
\hypersetup{
 pdfborderstyle=,allcolors=blue}

\makeatletter
\usepackage{iopams}
\usepackage{setstack}
\usepackage[normalem]{ulem}

\usepackage{color}

\newcommand{\dbar}{ \mathrm{ \mathchar'26\mkern-11.5 mu d}}

\def\rd{{\rm d}}
\def\vb{{\bf b}}

\def\vx{{\bf x}}
\def\mA{{\bf A}}
\def\mB{{\bf B}}

\def\mD{{\bf D}}

\def\mS{{\bf S}}
\def\vJ{{\bf J}}

\def\mQ{{\bf Q}}

\def\vX{{\bf X}}
\def\vY{{\bf Y}}

\def\vy{{\bf y}}
\def\vf{{\boldsymbol{f}}}

\def\vgamma{\mbox{\boldmath$\gamma$}}

\def\dbar{\thinspace {\mathchar'26 \mkern-12mu \rd}}

\makeatother

\begin{document}
\title{Potentials of Continuous Markov Processes and Random Perturbations}
\author{Ying-Jen Yang}
\ead{yangyj@uw.edu}
\address{Department of Applied Mathematics, University of Washington, Seattle,
98195, USA}
\author{Yu-Chen Cheng}
\ead{yuchench@uw.edu}
\address{Department of Applied Mathematics, University of Washington, Seattle,
98195, USA}
\begin{abstract}
With a scalar potential and a bivector potential, the  {vector field associated with the drift} of a diffusion
is decomposed into a generalized gradient field, a field perpendicular
to the gradient, and a divergence-free field. We give such decomposition
a probabilistic interpretation by introducing cycle velocity from
a bivectorial formalism of nonequilibrium thermodynamics. New understandings
on the mean rates of thermodynamic quantities are presented. Deterministic
dynamical system is further proven to admit a generalized gradient
form with the emerged potential as the Lyapunov function by the method
of random perturbations. 
\end{abstract}

\section{Introduction}

In mathematics, one can often gain a deeper understanding of an equation
when it is expressed in terms of its solutions. While this approach
might not sound relevant in engineering, where the goal is to \emph{find}
the solution(s), it has been extremely fruitful in theoretical science.
A case in point is to re-write the polynomial equation $x^{n}+a_{n-1}x^{n-1}+\cdots a_{1}x+a_{0}=0$
into $(x-x_{0})(x-x_{1})\cdots(x-x_{n})=0$. In a sense, one could
say the set of $n$ roots collectively \emph{defines} the algebraic
equation! Indeed, providing the setting for the existence and uniqueness
of such a \emph{re-writing} has become one of the most significant
chapters of mathematics \cite{berlinski_infinite_2008}.

In stochastic thermodynamics, the aforementioned philosophy translates
to rewriting dynamic equations in terms of their time-invariant solution,
the stationary probability density $\pi$. For equilibrium dynamics
with detailed balance, this has yielded Boltzmann's law that relates
the equilibrium distribution to a potential function, as well as the
celebrated {\em fluctuation-dissipation theorem}. For nonequilibrium
dynamics, the \emph{mesoscopic potential $\Phi$} as the negative
logarithm of $\pi$ has been discussed since the 1970s \cite{kubo_fluctuation_1973,graham_non-equilibrium_1983,nicolis_comment_1977}
up to the present day \cite{seifert_stochastic_2012,thompson_nonlinear_2016,yang_unified_2020}.
The introduction of $\Phi$ has led to a general force field decomposition
\cite{wang_potential_2008}, provided a notion of ``energy'' in
general nonequilibrium systems \cite{thompson_nonlinear_2016}, and
identified two sources of entropy production \cite{seifert_stochastic_2012,yang_unified_2020,ge_extended_2009,esposito_three_2010}.
In this {paper}, we build on the recently-revealed bivectorial structure
of nonequilibrium steady state (NESS) \cite{yang_bivectorial_2021}
to further decompose the force field in a physically-meaningful manner.

The fundamental roles of kinetic cycles of NESS in continuous Markov
processes without detailed balance lead to a bivectorial formalism
for stationary continuous processes \cite{yang_bivectorial_2021}.
In the present work, we introduce \emph{cycle velocity $\mathbf{Q}$}
based on the cycle flux $\mathbf{A}$ introduced in Ref. \cite{yang_bivectorial_2021}
and demonstrate its importance in the thermodynamics of continuous
Markov processes. Surprisingly, the resulting force decomposition
coincides with the decomposition extensively discussed by P. Ao \emph{et.
al.} \cite{ao_potential_2004,yin_existence_2006}. Our work shows
its generality and reveals its probabilistic origin with a novel cycle
interpretation. We further show that as the {vector field associated with the drift in} the diffusion
has a scalar potential $\Phi$ and a bivector potential $\mathbf{Q}$,
the probability flux has a scalar potential given by the ``\emph{free
energy}'' $F=\Phi-S$ and the cycle flux $p\mathbf{Q}$ as the bivector
potential where $p$ is the probability density and $S=-\ln p$ is
the Shannon entropy. Novel expressions on the mean rates of thermodynamic
quantities can then be derived.

In genernal, the mesoscopic potential $\Phi$ is not a Lyapunov function
of the underlying deterministic dynamics in a nonequilibrium diffusion.
However, in the small-noise limit, the locally-smooth \emph{macroscopic
potential} $\varphi$ emerges from the globally-smooth $\Phi$ and
is guaranteed to be the Lyapunov function of the deterministic dynamics
$\mathbf{x}'(t)=\mathbf{b}(\mathbf{x})$ \cite{freidlin_random_2012, qian_kinematic_2020}.
This provides us an orthogonal decomposition of the vector field $\mathbf{b}=-\mathbf{D}\nabla\varphi+\boldsymbol{\gamma}$
where $\boldsymbol{\gamma}\perp\nabla\varphi$ as shown by Freidlin
and Wentzell (FW) \cite{freidlin_random_2012}. It is further shown
in Ref. \cite{qian_kinematic_2020} that the decomposition is directly
related to the total entropy production decomposition in the small-noise
limit. In this {paper}, we show the general validity of a gradient
form of $\boldsymbol{\gamma}$, $\boldsymbol{\gamma}=-\mathbf{Q}\nabla\varphi$.
The bivectorial formalism further allows us to unify and compare the various random perturbations in Refs. \cite{ao_potential_2004,yin_existence_2006,freidlin_random_2012,qian_kinematic_2020}.
A continuous Markov process with a steady state has an energetics
given by the potential $\Phi$, the bivector potential $\mathbf{Q}$
and the diffusion matrix $\mathbf{D}$ from its thermodynamics, and
the energetics of a deterministic dynamics emerges in the small-noise,
thermodynamic limit.

\section{Two Representations of Continuous Markov Processes}

A continuous  {time-homogeneous Markov process\footnote{If the process is time inhomogeneous, our discussion can still be valid as long as the time dependent $\vb$ and $\mD$ still gives a unique invariant measure at each moment.} on $\mathbb{R}^{n}$ has two quite different
representations \cite{hong_representations_2020}. 
One is based on stochastic trajectories $\mathbf{X}_{t}$ with its probability
measure. The trajectories satisfy a stochastic differential equation
(SDE) 
\begin{equation}
\mathrm{d}\mathbf{X}_{t}=\left[\mathbf{b}(\mathbf{X}_{t})+\nabla\cdot\mathbf{D}(\mathbf{X}_{t})\right]\mathrm{d}t+\sqrt{2\mathbf{D}(\mathbf{X}_{t})}\mathrm{d}\mathbf{W}_{t}.\label{eq: SDE}
\end{equation}
The term $\mathbf{b}+\nabla\cdot\mathbf{D}$ is called the drift of the diffusion and $\mD$ is the diffusion matrix.
$\left(\nabla\cdot\mathbf{D}\right)_{j}=\sum_{i=1}^{n}\partial_{i}D_{ij}$
with $\partial_{i}$ denotes partial derivative with respect to $x_{i}$.
$\mathbf{W}_{t}$ is the $n$-D Brownian motion, and $\sqrt{2\mathbf{D}}$
is understood as  a matrix $\boldsymbol{\Gamma}$ such that $2\mathbf{D}=\boldsymbol{\Gamma}\boldsymbol{\Gamma}^{\mathsf{T}}$.
The other representation is based on
the (transition) probability density $p(\mathbf{x},t)$ with the Fokker-Planck
equation (FPE) 
\begin{equation}
\partial_{t}p(\vx,t)=-\nabla\cdot\left[\mathbf{b}(\mathbf{x})p(\vx,t)-\mathbf{D}(\mathbf{x})\nabla p(\vx,t)\right].\label{eq: FPE}
\end{equation} 
Correspondence between the two representations can be established
by Ito's calculus}\footnote{A trajectory $\vX_{t}$ satisfying the SDE has a unique representation
of the transitional probability density satisfying the FPE. This was
developed by the semigroup approaches \cite{feller_general_1954}.
However, if we start from the FPE alone and attempt to derive a unique
SDE representation from it, we would need the uniqueness of its semigroup
solution. The uniqueness requires quite restrictive conditions on
the drift and diffusion matrix. See Chap. 3 of Ref. \cite{jiang_mathematical_2004}
for how to relax the conditions, obtain the minimal semigroup solution,
and construct the corresponding continuous path $\vX_{t}$.}.

With Eq. \eref{eq: FPE} as the continuity equation of $p$, the
probability flux is given by 
\begin{equation}
\mathbf{J}(\mathbf{x},t)=\mathbf{b}(\mathbf{x})p(\mathbf{x},t)-\mathbf{D}(\mathbf{x})\nabla p(\mathbf{x},t).\label{eq: probability flux}
\end{equation}
This gives a notion of probability velocity as 
\begin{equation}
\mathbf{v}(\mathbf{x},t)=\frac{1}{p}\mathbf{J}=\mathbf{b}(\mathbf{x})+\mathbf{D}(\mathbf{x})\nabla S(\mathbf{x},t)\label{eq: prob velocity}
\end{equation}
where $S(\mathbf{x},t)=-\ln p(\mathbf{x},t)$ is the (stochastic)
Shannon entropy. Following Ref. \cite{lebowitz_gallavotti-cohen-type_1999},
the total heat dissipation in a infinitesimal time interval $\left(t,t+\mathrm{d}t\right)$
is given by 
\begin{equation}
\dbar\mathcal{Q}=\mathbf{D}^{-1}(\mathbf{X}_{t})\mathbf{b}(\mathbf{X}_{t})\circ\mathrm{d}\mathbf{X}_{t}\label{eq: heat dissipation functional}
\end{equation}
where $\circ$ indicates Stratonovich mid-point integration. {By interpreting Eq. \eref{eq: heat dissipation functional} as a heat dissipation, we have assumed that $\mD$ is invertible and that all the state variables $\vx$ and variables $\vb$ and $\mD$ are even-parity under time reversal.} The vector
field $\mathbf{D}^{-1}\mathbf{v}$ gives the \emph{thermodynamic}
\emph{force} of the (stochastic) total entropy production \cite{yang_unified_2020,yang_bivectorial_2021},
\label{dStot} 
\numparts
\begin{eqnarray}
\mathrm{\dbar}\mathcal{S}_{\mathrm{tot}} & =\mathrm{d}S+\dbar\mathcal{Q}\label{eq: total entropy production SDE}\\
 & =\partial_{t}S\left(\mathbf{X}_{t},t\right)\mathrm{d}t+\mathbf{D}^{-1}(\mathbf{X}_{t})\mathbf{v}(\mathbf{\mathbf{X}_{t}},t)\circ\mathrm{d}\mathbf{X}_{t}.\label{eq: dStot partial S D-1 v}
\end{eqnarray}
\endnumparts
The decomposition in Eq. \eref{eq: prob velocity} reflects the two
origins of the thermodynamic force $\mathbf{D}^{-1}\mathbf{v}$: the
force $\mathbf{D}^{-1}\mathbf{b}$ of total heat dissipation \cite{lebowitz_gallavotti-cohen-type_1999}
and the entropic force $\nabla S$ in the entropy change $\mathrm{d}S=\partial_{t}S\mathrm{d}t+\nabla S\circ\mathrm{d}\mathbf{X}_{t}$.

\section{Decomposition of Diffusion}

\subsection{Entropy Production Decomposition}

We assume the system has a steady state with an invariant density
$\pi(\mathbf{x})$. The divergence-free stationary probability flux
is then given by 
\begin{equation}
\mathbf{J}^{*}(\mathbf{x})=\mathbf{b}(\mathbf{x})\pi(\mathbf{x})-\mathbf{D}(\mathbf{x})\nabla\pi(\mathbf{x}).\label{eq: stationary prob flux}
\end{equation}
And, the stationary probability velocity is 
\begin{equation}
\mathbf{v}^{*}(\mathbf{x})=\mathbf{b}(\mathbf{x})+\mathbf{D}(\mathbf{x})\nabla\Phi(\mathbf{x})\label{eq: stationary prob v}
\end{equation}
where $\Phi(\mathbf{x})=-\ln\pi(\mathbf{x})$. We shall call the potential
$\Phi$ the \emph{mesoscopic potential} to later distinguish it with
the macroscopic potential that emerges from it in the small-noise
limit. Eq. \eref{eq: stationary prob v} leads to a decomposition
of $\mathbf{D}^{-1}\mathbf{b}$ \cite{wang_potential_2008}, 
\begin{equation}
\mathbf{D}^{-1}\mathbf{b}(\mathbf{x})=-\nabla\Phi(\mathbf{x})+\mathbf{D}^{-1}\mathbf{v}^{*}(\mathbf{x}),\label{eq:  phi v* decomposition  of b}
\end{equation}
and a decomposition of the thermodynamic force 
\begin{equation}
\mathbf{D}^{-1}(\mathbf{x})\mathbf{v}(\mathbf{x},t)=-\nabla F(\mathbf{x},t)+\mathbf{D}^{-1}(\mathbf{x})\mathbf{v}^{*}(\mathbf{x})\label{eq: decomposition of Dinv v}
\end{equation}
where $F(\mathbf{x},t)=\Phi(\mathbf{x})-S(\mathbf{x},t)$ is understood
as the ``free energy'' in nonequilibrium systems \cite{thompson_nonlinear_2016, qian_relative_2001}.
This decomposition in Eq. \eref{eq: decomposition of Dinv v} corresponds
to the celebrated (stochastic) total entropy production decomposition
$\thinspace\dbar\mathcal{S}_{\mathrm{tot}}=-\mathrm{d}F+\thinspace\dbar\mathcal{Q}_{\mathrm{hk}}$
where $\mathcal{Q}_{\mathrm{hk}}$ is the housekeeping heat dissipation.
See Ref. \cite{yang_unified_2020} and the references within for a
recent synthesis. Here we note a recent study showing rigorously how
the housekeeping heat dissipation in a compact, driven process can
be mapped to the energy dissipation of a lifted, detailed-balanced
process \cite{wang_mathematical_2020}.

The decomposition in Eq. \eref{eq: decomposition of Dinv v} can
be interpreted as a decomposition of the FPE generator \cite{qian_decomposition_2013,qian_zeroth_2014,qian_thermodynamics_2015},
\begin{eqnarray}
\partial_{t}p & =-\nabla\cdot\left(p\mathbf{v}^{*}\right)-\nabla\cdot\left[p\left(-\mathbf{D}\nabla F\right)\right].\label{eq: FPE decomposition}
\end{eqnarray}
The former on the right-hand-side of Eq. \eref{eq: FPE decomposition}
corresponds to a Liouville equation 
\begin{equation}
\partial_{t}p=-\nabla\cdot\left(p\mathbf{v}^{*}\right)\label{eq: Liouville}
\end{equation}
of a measure-preserving deterministic dynamical system $\mathbf{x}'(t)=\mathbf{v}^{*}(\mathbf{x})$
with $e^{-\Phi}$ as an invariant measure. The latter corresponds
to a detailed-balanced diffusion process with the same invariant density
$e^{-\Phi}$ described by 
\begin{eqnarray}
\partial_{t}p & =-\nabla\cdot\left[p\left(-\mathbf{D}\nabla F\right)\right].\label{eq: DB FPE}
\end{eqnarray}
That is, every diffusion process can be regarded as a deterministic
dynamical system coupled with the ``randomly-damping'', detailed-balanced
environment \cite{qian_zeroth_2014,qian_thermodynamics_2015}. {We note that $e^{-\Phi}$ is the invariant measure \emph{before} and
\emph{after} the coupling between Eq. \eref{eq: Liouville} and Eq. \eref{eq: DB FPE}. This is considered as a generalization of \emph{the
zeroth law of thermodynamics} in Ref. \cite{qian_zeroth_2014}.}

\subsection{Bivectorial Decomposition with Cycle Velocity}

The divergent-free stationary current $\mathbf{J}^{*}$ can be furthered
expressed as the $n$-D ``curl'' of a \emph{bivector} $\mathbf{A}$,
an anti-symmetric matrix that represents \emph{cycle flux} \cite{yang_bivectorial_2021},
\begin{equation}
\mathbf{J}^{*}=\nabla\times\mathbf{A}\label{eq: bivector potential}
\end{equation}
where $\left(\nabla\times\mathbf{A}\right)_{i}=\sum_{j=1}^{n}\partial_{j}A_{ij}$. 
{While solving an explicit form of $\mA$ requires a non-trivial calculation on solving Eq. \eref{eq: bivector potential}, this expression actually allows us to reveal more structure of the system as shown below.}
The stationary probability velocity $\mathbf{v}^{*}$ can then be
expressed as 
\begin{eqnarray}
\mathbf{v}^{*} & =e^{\Phi}\nabla\times\mathbf{A}=\nabla\times\left(e^{\Phi}\mathbf{A}\right)-\mathbf{A}\nabla e^{\Phi}.\label{eq: v* in terms of A}
\end{eqnarray}
We then introduce\emph{ }the\emph{ cycle} \emph{velocity} 
\begin{equation}
\mathbf{Q}=\frac{1}{\pi}\mathbf{A}=e^{\Phi}\mathbf{A},\label{eq: cycle velocity}
\end{equation}
which is also a \emph{bivector}. The stationary velocity $\mathbf{v}^{*}$
then has the decomposition 
\begin{equation}
\mathbf{v}^{*}=-\mathbf{Q}\nabla\Phi+\nabla\times\mathbf{Q}\label{eq: v* decomposition in terms of Q}
\end{equation}
where the former is perpendicular to $\nabla\Phi$ and the latter
is divergence-free. This type of rewriting was first proposed mathematically
by Graham \cite{graham_covariant_1977} and recently by \cite{ding_covariant_2020}.
The vector field $\mathbf{b}$ now has a decomposition in terms of
$\Phi,$ $\mathbf{Q}$, and $\mathbf{D}$, 
\begin{equation}
\mathbf{b}=-\mathbf{D}\nabla\Phi-\mathbf{Q}\nabla\Phi+\nabla\times\mathbf{Q},\label{eq: b decomposition}
\end{equation}
This decomposition actually has been extensively discussed by P. Ao
\emph{el. al.} in the past decades \cite{ao_potential_2004,yin_existence_2006,  shi_relation_2012}.
Here, we show its generality and reveal a novel probabilistic origin
with a cycle interpretation. We also derive the following novel decomposition
of the probability flux 
\begin{equation}
\mathbf{J}=p\mathbf{v}=-p\mathbf{D}\nabla F-p\mathbf{Q}\nabla F+\nabla\times\left(p\mathbf{Q}\right).\label{eq: J decomposition}
\end{equation}
In a word, $(\Phi,\mathbf{Q})$ are the scalar and bivector potentials
of the {vector field} $\mathbf{b}$, and $(F,p\mathbf{Q})$ are the scalar
and bivector potentials of the flux $\mathbf{J}$!

\subsection{Mean Rate Decomposition of Thermodynamic Quantities.}

The above probability flux decomposition in Eq. \eref{eq: J decomposition}
leads to a new understanding on the mean rates of various thermodynamic
quantities. Following Ref. \cite{yang_bivectorial_2021}, various
mean rates can be derived by considering a general work-like quantity
$\mathcal{W}$ whose infinitesimal change satisfies $\thinspace\dbar\mathcal{W}=\boldsymbol{f}(\mathbf{X}_{t},t)\circ\mathrm{d}\mathbf{X}_{t}$
with a force field $\boldsymbol{f}(\mathbf{x},t)$. With $\mathbb{E}\left[\cdot\right]$
denoting expectation with respect to $p(\mathbf{x},t)$, the mean
rate of $\mathcal{W}$ has the following decomposition 
\begin{eqnarray}
\dot{\mathcal{W}}=\frac{\mathbb{E}[\dbar\mathcal{W}]}{\mathrm{d}t}=\mathbb{E}\left[\left(-\mathbf{D}\nabla F\right)\cdot\boldsymbol{f}\right]+\mathbb{E}\left[\mathbf{v}^{*}\cdot\boldsymbol{f}\right].\label{eq: mean rate of work decomposition}
\end{eqnarray}
The second term, by integration by part, can be rewritten as 
\begin{eqnarray}
\mathbb{E}\left[\mathbf{v}^{*}\cdot\boldsymbol{f}\right] & =\mathbb{E}\left[\mathbf{Q}\cdot\nabla\wedge\boldsymbol{f}\right]+\mathbb{E}\left[\mathbf{Q}\cdot\boldsymbol{f}\wedge\left(-\nabla F\right)\right].\label{eq: f v* decomposition}
\end{eqnarray}
Both wedge terms are bivectors with components $\left(\mathbf{u}\wedge\mathbf{w}\right)_{ij}=u_{i}w_{j}-u_{j}w_{i}$
for $1\le i<j\le n$. The scalar products in Eq. \eref{eq: f v* decomposition}
between two bivectors are the half of the {Frobenius products between
matrices, $\mA \cdot \mB =\sum_{i<j} A_{ij} B_{ij} = \frac{1}{2}\sum_{i,j} A_{ij} B_{ij}$ \cite{yang_bivectorial_2021}.} Since $\Vert\mathbf{u}\wedge\mathbf{w}\Vert=\sqrt{\left(\mathbf{u}\wedge\mathbf{w}\right)\cdot\left(\mathbf{u}\wedge\mathbf{w}\right)}$
is the area of the parallelogram of $\mathbf{u}$ and $\mathbf{w}$
in $\mathbb{R}^{n},$ a (simple) bivector $\mathbf{u}\wedge\mathbf{w}$
can be understood as a generalized ``signed'' area of in $\mathbb{R}^{n}$.
See supplemental material for a brief introduction.

{We note that the notion of curl in $\mathbb{R}^3$ is generalized to $\mathbb{R}^n$ by two different operations: $\nabla \times$ and $\nabla \wedge$ as shown in Ref. \cite{yang_bivectorial_2021}. 
The former maps a bivector to a divergence free vector field as shown in Eq. \eref{eq: bivector potential}, while the latter maps a vector field $\vf$ to a bivector $\nabla \wedge \vf$ representing the vorticity of the vector field.} 

The decomposition in Eq. \eref{eq: f v* decomposition} shows the
fundamental roles of cycles in nonequilibrium thermodynamics with
nonzero $\mathbf{v}^{*}$. Both terms in Eq. \eref{eq: f v* decomposition}
are cyclic averages of bivectors, both averaged over the ``cycle
flux'' $pQ_{ij}$ in each infinitesimal plane $\mathrm{d}x_{i}\wedge\mathrm{d}x_{j}$.
This is an extension to Ref. \cite{yang_bivectorial_2021} where cycle
flux $\mathbf{A}=\pi\mathbf{Q}$ was first introduced in NESS. We
note that as the system approaches NESS as $t\rightarrow\infty$,
the term $\mathbb{E}\left[\mathbf{Q}\cdot\nabla\wedge\boldsymbol{f}\right]$
persists whereas $\mathbb{E}\left[\mathbf{Q}\cdot\boldsymbol{f}\wedge\left(-\nabla F\right)\right]\rightarrow0$
since $F\rightarrow0$.

The two terms in Eq. \eref{eq: f v* decomposition} have the following
physical interpretations. With $\nabla\wedge\boldsymbol{f}$ being
the $n$-D ``curl'' of vectors, its cyclic average $\mathbb{E}\left[\mathbf{Q}\cdot\nabla\wedge\boldsymbol{f}\right]$
is the mean circulation of the force $\boldsymbol{f}$. Hence, if
a force is a gradient vector field $\boldsymbol{f}=-\nabla U$, then
$\nabla\wedge\left(-\nabla U\right)=\mathbf{0}$. This implies $\mathbb{E}\left[\mathbf{Q}\cdot\nabla\wedge\boldsymbol{f}\right]$
would be zero in the mean rate of state observables: $S,$ $\Phi$,
and $F$. On the other hand, the wedge product $\boldsymbol{f}\wedge\left(-\nabla F\right)$
is the generalized ``signed'' area spanned by the two vectors $\boldsymbol{f}$
and $-\nabla F$. $\mathbb{E}\left[\mathbf{Q}\cdot\boldsymbol{f}\wedge\left(-\nabla F\right)\right]$
is thus a ``torque-like'' quantity representing the mean area between
the force $\boldsymbol{f}$ and $-\nabla F$ averaged over all its
planar components. It would be zero when $\boldsymbol{f}$ is parallel
to $-\nabla F$.

Eq. \eref{eq: f v* decomposition} shows us that the average orthogonality
between $\nabla F$ and $\mathbf{v}^{*}$ discussed in Ref. \cite{yang_bivectorial_2021}
is due to both $\nabla F$ being a gradient field and parallel to
$-\nabla F$. The mean rate of free energy then has the expression
\begin{equation}
\dot{F}=\mathbb{E}\left[-\left(\nabla F\right)\cdot\mathbf{D}\nabla F\right]\le0.\label{eq: free energy dissipation rate}
\end{equation}
It is a purely ``gradient-descending'' term with no cyclic contribution,
reflecting the detailed-balanced dynamics of Eq. \eref{eq: DB FPE}
hidden behind. The housekeeping heat dissipation rate, on the other
hand, has the following new expression 
\begin{eqnarray}
\dot{\mathcal{Q}}_{\mathrm{hk}}= & \mathbb{E}\left[\mathbf{Q}\cdot\nabla\wedge \left(\mathbf{D}^{-1}\mathbf{b}\right)\right]-\mathbb{E}\left[\mathbf{Q}\cdot \left(\mathbf{D}^{-1}\mathbf{v}^{*}\right)\wedge\nabla F\right]\label{eq: Qhk rate}
\end{eqnarray}
where the first term is the average cycle affinity \cite{yang_bivectorial_2021}.
Eq. \eref{eq: Qhk rate} is purely cyclic, reflecting the measure-preserving
deterministic dynamics of Eq. \eref{eq: Liouville} hidden behind.

The mean rates of entropy and mesoscopic potential now have the following
new expressions: \label{Sdot and Phidot} 
\numparts
\begin{eqnarray}
\dot{S}= & +\underset{\ge0}{\underbrace{\mathbb{E}\left[\nabla S\cdot\mathbf{D}\nabla S\right]}}-r+w\label{eq: Sdot}\\
\dot{\Phi}= & -\underset{\ge0}{\underbrace{\mathbb{E}\left[\nabla\Phi\cdot\mathbf{D}\nabla\Phi\right]}}+r+w\label{eq: Phi dot}
\end{eqnarray}
\endnumparts
where $w=\mathbb{E}\left[\mathbf{Q}\cdot\left(\nabla\Phi\wedge\nabla S\right)\right]$
denotes a wedge product term and $r=\mathbb{E}\left[\nabla S\cdot\mathbf{D}\nabla\Phi\right]$
denotes a scalar product term. Besides the source/sink terms, the two rates
have two common contributions: a ``curl'' $w$ measuring the perpendicularity
between $\nabla\Phi$ and $\nabla S$ and an inner product $r$. As
the system approaches NESS, $S\rightarrow\Phi$ and the wedge product
term $w\rightarrow0$. Both $\mathbb{E}\left[\nabla S\cdot\mathbf{D}\nabla S\right]$
and $r$ converge to $\mathbb{E}\left[\nabla\Phi\cdot\mathbf{D}\nabla\Phi\right]$,
canceling each other out. 

\section{Mesoscopic Potential and the Emerged Macroscopic Potential}

\subsection{Mesoscopic Potential and the Maxwell-Boltzmann Equilibrium}

In equilibrium physics, a ``potential function'' has many important
features. Among those stand the following two prominent ones \cite{huang_processes_2017}:
1) It is related to the equilibrium probability distribution via the
Maxwell-Boltzmann (M-B) distribution; 2) It is a Lyapunov function
of the underlying deterministic dynamics. The mesoscopic potential
$\Phi$ satisfies the first feature by it's definition and achieved
the second feature by the equilibrium condition $\mathbf{v}^{*}=\mathbf{0}$:
it is the Lyapunov function of $\mathbf{x}'(t)=\mathbf{b}(\mathbf{x})=-\mathbf{D}(\mathbf{x})\nabla\Phi(\mathbf{x})$,{\emph{i.e.} $\frac{\rd}{\rd t} \Phi(\vx(t))\le 0$ the dynamics is going downhill of $\Phi$}.
{We would call $-\mD \nabla \Phi$ a \emph{generalized gradient term} throughout this paper. This naming is to emphasize that it shares many features with a typical gradient term. In particular, the dynamics $\mathbf{x}'(t)=-\mathbf{D}(\mathbf{x})\nabla\Phi(\mathbf{x})$ can not have any cyclic trajectory, similar to a gradient-descending dynamics $\mathbf{x}'(t)=-\nabla\Phi(\mathbf{x})$. In fact, the dynamics $\mathbf{x}'(t)=-\mathbf{D}(\mathbf{x})\nabla\Phi(\mathbf{x})$ is the gradient-descending dynamics on a manifold with metric tensor given by $\mD^{-1}$.} 

For general nonequilibrium systems, however, $\Phi$ is \emph{not}
always the Lyapunov function of the deterministic dynamics $\mathbf{x}'(t)=\mathbf{b}(\mathbf{x})$.
The necessary and sufficient condition of the Lyapunov property is
\begin{equation}
\nabla\Phi\cdot\mathbf{v}^{*}\le\nabla\Phi\cdot\mathbf{D}\nabla\Phi,\label{eq: Lyapunov condition: perpendicular}
\end{equation}
which can be further rewritten as 
\begin{equation}
\nabla\cdot\mathbf{v}^{*} \le\nabla\Phi\cdot\mathbf{D}\nabla\Phi\label{eq: Lyapunov condition: div free}
\end{equation}
by using the stationary FPE 
\begin{equation}
\nabla\cdot\left(\pi\mathbf{v}^{*}\right)=0\Leftrightarrow\mathbf{v}^{*}\cdot\nabla\Phi=\nabla\cdot\mathbf{v}^{*}.\label{eq: stationary FPE and perpendicular, div}
\end{equation}
In particular, a sufficient condition of the Lyapunov property of
$\Phi$ is 
\begin{equation}
\nabla\cdot\mathbf{v}^{*}=\mathbf{v}^{*}\cdot\nabla\Phi=0.\label{eq: M-B equilibrium condition}
\end{equation}
Nonequilibrium systems with Eq. (\ref{eq: M-B equilibrium condition})
is said to admit \emph{Maxwell-Boltzmann (M-B)} \emph{equilibrium} in Ref. \cite{qian_thermodynamics_2015}. In such systems, $\Phi$ is a Lyapunov
function of $\mathbf{x}'(t)=\mathbf{b}(\mathbf{x})$, and the vector
field $\mathbf{b}$ can be decomposed into a generalized gradient
term $-\mathbf{D}\nabla\Phi$ and a \emph{divergent-free} term $\mathbf{v}^{*}$,
akin to the Helmholtz decomposition in $\mathbb{R}^{3}$. Furthermore,
with M-B equilibrium, the deterministic dynamics $\mathbf{x}'(t)=\mathbf{v}^{*}(\mathbf{x})$
corresponding to the Liouville equation in Eq. \eref{eq: Liouville}
has a divergent-free vector field $\nabla\cdot\mathbf{v}^{*}=0$ and
a conserved quantity $\Phi(\mathbf{x})$. It is a generalization of
Hamiltonian systems with energy $\Phi$ \cite{qian_zeroth_2014,qian_thermodynamics_2015}.

In the small-noise limit where the diffusion becomes a random perturbation
to the deterministic dynamical system $\mathbf{x}'(t)=\mathbf{b}(\mathbf{x})$,
a macroscopic potential $\varphi(\mathbf{x})$ emerges from the mesoscopic
potential $\Phi$ and is guaranteed to be a Lyapunov function of the
deterministic dynamics \cite{freidlin_random_2012,qian_kinematic_2020}.
In fact, in the small-noise limit, the decomposition $\mathbf{b}=-\mathbf{D}\nabla\Phi+\mathbf{v}^{*}$
becomes a perpendicular decomposition 
\begin{equation}
\mathbf{b}=-\mathbf{D}\nabla\varphi+\boldsymbol{\gamma}\label{eq: varphi gamma decomposition}
\end{equation}
{where $\nabla \varphi \perp \boldsymbol{\gamma}$. Eq. \eref{eq: varphi gamma decomposition}} is connected to the decomposition of total entropy production
rate by Ref. \cite{qian_kinematic_2020}. We shall see in the next
section that the cycle velocity decomposition in Eq. \eref{eq: b decomposition}
further reveals that $\boldsymbol{\gamma}=-\mathbf{Q}\nabla\varphi$.

\subsection{Emergent Macroscopic Potential and Decomposition of Dynamical Systems}

Let us characterize the small-noise limit with a parameter $\epsilon$.
From now on, we denote $\epsilon$-dependence of variables with a
subscript $\epsilon$, e.g. $\mathbf{b}_{\epsilon}(\mathbf{x})$.
A variable without subscript $\epsilon$ is $\epsilon$-independent.
With a {vector field} $\mathbf{b}_{\epsilon}$ and a diffusion matrix
$\mathbf{D}_{\epsilon}$, the resulting $\Phi_{\epsilon}$ and $\mathbf{Q}_{\epsilon}$
are both $\epsilon$-dependent in general. Following Ref. \cite{qian_kinematic_2020},
we start with a deterministic dynamics $\mathbf{x}'(t)=\mathbf{b}(\mathbf{x})$
and impose a random perturbation to get a diffusion described by the
FPE, 
\begin{equation}
\partial_{t}p=\nabla\cdot\left[\epsilon\mathbf{D}\nabla p-\mathbf{b}p\right].\label{eq: FPE of random perturbation}
\end{equation}
and the corresponding SDE, 
\begin{equation}
\mathrm{d}\mathbf{X}_{t}=\left(\mathbf{b}+\epsilon\nabla\cdot\mathbf{D}\right)\mathrm{d}t+\sqrt{2\epsilon\mathbf{D}}\mathrm{d}\mathbf{W}_{t}.\label{eq: SDE of random perturbation}
\end{equation}
By Eq. \eref{eq: b decomposition}, we have a generally-valid decomposition
\begin{equation}
\mathbf{b}=-\epsilon\mathbf{D}\nabla\Phi_{\epsilon}-\mathbf{Q}_{\epsilon}\nabla\Phi_{\epsilon}+\nabla\times\mathbf{Q}_{\epsilon}.\label{eq: b in epsilon decomposition}
\end{equation}

By applying the WKB ansatz on the invariant density {$\pi_{\epsilon}=\omega(\vx)e^{-\frac{\varphi(\vx)}{\epsilon}+a\ln\epsilon+O(\epsilon)}$},
we have an asymptotic series of $\Phi_{\epsilon}$ \cite{qian_kinematic_2020},
\begin{equation}
\Phi_{\epsilon}(\mathbf{x})=\frac{\varphi(\mathbf{x})}{\epsilon}-\ln\omega(\mathbf{x})-a\ln\epsilon+O(\epsilon).\label{eq: Phi asymptotic}
\end{equation}
Equations for $\varphi$ can be obtained by plugging this back to
the stationary FPE, $\nabla\cdot\mathbf{J}_{\epsilon}^{*}=0$. One
gets an Hamilton-Jacobi equation (HJE) of $\varphi$, $0=-\boldsymbol{\gamma}\cdot\nabla\varphi$,
which leads to the aforementioned decomposition of the vector field
$\mathbf{b}$, 
\begin{equation}
\mathbf{b}(\mathbf{x})=-\mathbf{D}(\mathbf{x})\nabla\varphi(\mathbf{x})+\boldsymbol{\gamma}(\mathbf{x})\label{eq: b varphi gamma decomposition}
\end{equation}
with $\varphi$ guaranteed to be the Lyapunov function of $\mathbf{x}'(t)=\mathbf{b}(\mathbf{x})$
\cite{qian_kinematic_2020}. Now by plugging Eq. \eref{eq: Phi asymptotic}
in to Eq. \eref{eq: b in epsilon decomposition}, one sees that $\mathbf{Q}_{\epsilon}=O(\epsilon)$
and, by collecting the leading order terms, 
\begin{equation}
\mathbf{b}(\mathbf{x})=-\mathbf{D}\left(\mathbf{x}\right)\nabla\varphi\left(\mathbf{x}\right)-\mathbf{Q}\left(\mathbf{x}\right)\nabla\varphi\left(\mathbf{x}\right)\label{eq: D Q decomposition of b}
\end{equation}
where $\mathbf{Q}\left(\mathbf{x}\right)=\lim_{\epsilon\rightarrow0}\mathbf{Q}_{\epsilon}/\epsilon$.
Therefore, $\boldsymbol{\gamma}$ is directly related to the bivectorial
cycle velocity, 
\begin{equation}
\boldsymbol{\gamma}\left(\mathbf{x}\right)=-\mathbf{Q}\left(\mathbf{x}\right)\nabla\varphi\left(\mathbf{x}\right)\label{eq: gamma =00003D00003D -Q del varphi}
\end{equation}
which matches the result in Ref. \cite{qian_kinematic_2020} conceptually
that $\boldsymbol{\gamma}$ corresponds to the macroscopic housekeeping
dissipation rate, with cycles as fundamental units \cite{yang_bivectorial_2021}.
This shows that a dynamical system has a generally-valid decomposition
in a gradient form, 
\begin{equation}
\mathbf{x}'(t)=-\left(\mathbf{D}\left(\mathbf{x}\right)+\mathbf{Q}\left(\mathbf{x}\right)\right)\nabla\varphi\left(\mathbf{x}\right).\label{eq: x' =00003D00003D -(D+Q) del varphi}
\end{equation}
We note that Ao \emph{et. al.} have discussed the decomposition of
the same form in the past \cite{ao_potential_2004, zhu_limit_2006}.
Here, we link the potential in the decomposition to the quasipotential
in FW's theory and prove its generality for systems whose random perturbation
admits an invariant distribution. We note that $\varphi$ in general
can have nonsmoothness, for example\emph{,} on the separatrix between
basins of attraction \cite{freidlin_random_2012,qian_kinematic_2020,zhou_construction_2016}.
However, dynamics exactly on the separatrix is usually less relevant
in dynamical systems.

\subsection{Random Perturbations organized by the Bivectorial Formalism}

Several types of random perturbations have been considered in the
past literature \cite{ao_potential_2004,yin_existence_2006,freidlin_random_2012, qian_kinematic_2020,zhou_construction_2016}.
{The physical meanings of these various types of random perturbations can be found in their specific applications \cite{nolting_balls_2015,tang_potential_2017,cheng_stochastic_2021}. 
Here, we compare them by our bivectorial formalism, with focus on the various potentials that emerged in the small-noise limit of these different perturbations. In general, the randomly perturbed system is described by the following SDE \begin{equation} \mathrm{d}\mathbf{X}_{t}=\left(\mathbf{b}_{\epsilon}+\epsilon\nabla\cdot\mathbf{D}\right)\mathrm{d}t+\sqrt{2\epsilon\mathbf{D}}\mathrm{d}\mathbf{W}_{t}.\label{eq: SDE of random perturbation Sec 4.3}\end{equation}
and the following FPE
\begin{equation}
\partial_{t}p_{\epsilon}=\nabla\cdot\left[\epsilon\mathbf{D}\nabla p_{\epsilon}-\mathbf{b}_{\epsilon}p_{\epsilon}\right].\label{eq: FPE of random perturbation Sec 4.3}
\end{equation} 
In particular, we focus on three types of perturbation.} The perturbation we have considered so far corresponds to the case
where the {vector field} is $\epsilon$-independent $\mathbf{b}_{\epsilon}=\mathbf{b}$
\cite{qian_kinematic_2020}. The random perturbation considered by
Freidlin and Wentzell (FW) is with $\mathbf{b}_{\epsilon}=\mathbf{b}+\epsilon\left(-\nabla\cdot\mathbf{D}\right)$ \cite{freidlin_random_2012}. 
In Ref. \cite{yin_existence_2006},
Yin and Ao started from a 2nd order Klein-Kramer equation and took
a zero-mass limit to arrive {at} a random perturbation with scaling $\mathbf{Q}_{\epsilon}=\epsilon\mathbf{Q}$
and $\Phi_{\epsilon}=\phi/\epsilon$, leading to $\mathbf{b}_{\epsilon}=\mathbf{-\mathbf{D}\nabla\phi-\mathbf{Q}\nabla\phi+\epsilon\left(\nabla\times\mathbf{Q}\right)}.$
We see that the three perturbations correspond to different $\mathbf{b}_{1}$
in $\mathbf{b}_{\epsilon}=\mathbf{b}+\epsilon\mathbf{b}_{1}$.

By again using the WKB ansatz on the stationary distribution and the
stationary FPE, one can obtain the equations for $\varphi$ and $\omega$,\label{PDEs for varphi and omega}
\numparts
\begin{eqnarray}
0 & =-\boldsymbol{\gamma}\cdot\nabla\varphi\label{eq: HJE-1}\\
\nabla\cdot\left(\omega\boldsymbol{\gamma}\right) & =-\nabla\varphi\cdot\left[\mathbf{D}\nabla\omega-\mathbf{b}_{1}\right],\label{eq: omega PDE-1}
\end{eqnarray}
\endnumparts
where $\boldsymbol{\gamma}(\mathbf{x})=\mathbf{b}(\mathbf{x})+\mathbf{D}(\mathbf{x})\nabla\varphi(\mathbf{x})$.
That is, the potentials $\varphi$ from perturbations with different
$\mathbf{b}_{1}$ all satisfy the same HJE.

However, this alone does not tell us in which domains the three potentials
have the same shape. To answer that relies on the unique orthogonal
decomposition theorem by FW \cite{freidlin_random_2012}.
With respect to a stable fixed point $\mathbf{x}_{0}$ of $\mathbf{x}'(t)=\mathbf{b}(\mathbf{x})$,
FW defined quasipotential as the minimization of an action functional
$\mathcal{S}_{0,T}\left(\xi\right)$ with respect to $\mathbf{x}_{0}$,
\begin{equation}
\psi(\vx;\vx_{0}):=\inf_{T>0}\inf_{\xi\in\Xi_{T}}\{\mathcal{S}_{0,T}(\xi):\xi_{0}=\vx_{0},\xi_{T}=\vx\}\label{quasipotential}
\end{equation}
where $\Xi_{T}$ is the set of smooth paths on the interval $[0,T]$,
and $\mathcal{S}_{0,T}(\xi)=\frac{1}{4}\int_{0}^{T}[\dot{\xi_{s}}-\vb(\xi_{s})]\mD^{-1}(\xi_{s})[\dot{\xi_{s}}-\vb(\xi_{s})]\rd s$
is the action functional \cite{freidlin_random_2012}. The theorem
then states that the $\varphi$ in any orthogonal decomposition of
$\mathbf{b}(\mathbf{x})$ as Eq. \eref{eq: b varphi gamma decomposition},
no matter from which random perturbation, actually have the same shape
as the quasipotential $\psi\left(\mathbf{x};\mathbf{x}_{0}\right)$
in the domain where $\varphi\left(\mathbf{x}\right)>\varphi\left(\mathbf{x}_{0}\right)$,
$\varphi(\mathbf{x})$ is continuously differentiable and $\nabla\varphi\left(\mathbf{x}\right)\neq\mathbf{0}$.
We note that the quasipotential $\psi$ has a representation in terms
of the conditional probability density \cite{freidlin_random_2012,zhou_construction_2016}
\begin{equation}
\psi(\vx,\vx_{0})=-\lim_{t\rightarrow\infty}\lim_{\epsilon\rightarrow0}\epsilon\ln p_{\epsilon}(\vx,t|\vx_{0},0).\label{eq: quasi potential limit-1}
\end{equation}
The macroscopic potential $\varphi$, by its definition, has a different
order of limits, 
\begin{equation}
\varphi(\vx) =-\lim_{\epsilon\rightarrow0}\lim_{t\rightarrow\infty}\epsilon\ln p_{\epsilon}(\vx,t|\vx_{0},0).\label{eq:Ao as limit-1}
\end{equation}
The fact that they have the same shape locally is nontrivial.

In the domains where $\varphi(\mathbf{x})$ from the three perturbations
match, the leading order differences of the three $\Phi_{\epsilon}$
are in their prefactors $\omega(\mathbf{x})$, which satisfy different
Eq. \eref{eq: omega PDE-1} with different $\mathbf{b}_{1}$. In
Ao's perturbation, we have \textbf{$\boldsymbol{\gamma}(\mathbf{x})=-\mathbf{Q}(\mathbf{x})\nabla\phi(\mathbf{x})$}
and $\varphi(\mathbf{x})=\phi(\mathbf{x})$. Therefore, Eq. \eref{eq: omega PDE-1}
in Ao's perturbation becomes 
\begin{equation}
\nabla\omega(\vx)\cdot\left[\mD(\vx)-\mQ(\vx)\right]\nabla\varphi(\vx)=0.\label{eq: omega Ao}
\end{equation}
Ao's perturbation corresponds to the particular solution $\omega=1$
\footnote{The choice of $\omega=1$ is a particular solution of Eq. \eref{eq: omega Ao}
with specific boundary conditions and it is independent of the divergence
of $\boldsymbol{\gamma}(\vx)$. If we started with $\Phi_{\epsilon}(\mathbf{x})=\frac{\phi(\mathbf{x})}{\epsilon}-\ln\omega(\mathbf{x})$
in the perturbation where the non-uniform $\omega(x)$ satisfies Eq.
\eref{eq: omega Ao}, this then generalizes Ao's perturbation. Nice
properties of $\phi$ such as the global smoothness and Lyapunov function
are still preserved in this generalization.}. In comparison, $\omega$ is non-uniform in general in the other
two perturbations. In fact, $\omega=1$ is a sufficient condition
for $\nabla\cdot\boldsymbol{\gamma}=0$ in \cite{qian_kinematic_2020}
but not in Ao's perturbation.

\section{Conclusion and Discussion}

In this study, we extend the decomposition of continuous Markov process
associated with the total entropy production decomposition by introducing
the bivectorial cycle velocity. Hidden structures in the mean rate
of thermodynamic quantities are revealed. Further in the small noise
thermodynamic limit, the emergent dynamical system is shown to admit
a generalizated gradient form. Differences among three random perturbations
are further discussed with the bivectorial formalism.

The introduction of cycle velocity organizes the two notions of equilibrium
in thermodynamics: 1) Equilibrium in classical thermodynamics is detailed
balanced with zero stationary velocity $\mathbf{v}^{*}=\mathbf{0}$
and zero housekeeping heat dissipation $\mathcal{Q}_{\mathrm{hk}}=0$.
The cycle velocity satisfies $\nabla\times\mathbf{Q}=\mathbf{Q}\nabla\Phi$
in equilibrium. 2) The notation of M-B equilibrium introduced by Qian
\cite{qian_zeroth_2014,qian_thermodynamics_2015} is an extension
of the notion of equilibrium, from $\mathbf{v}^{*}=\mathbf{0}$ to
$\mathbf{v}^{*}\perp\nabla\Phi$. The orthoganality $\mathbf{v}^{*}\perp\nabla\Phi$
is equivalent to $\left(\nabla\times\mathbf{Q}\right)\bot\nabla\Phi$
in the bivectorial formalism since $\nabla\Phi\cdot\mathbf{Q}\nabla\Phi=0$.
Therefore, all systems with constant cycle velocity $\mathbf{Q}$
are in M-B equilibrium, e.g. the Ornstein-Uhlenbeck process \cite{kwon_structure_2005}.
For systems with M-B equilibrium, the potential $\Phi$ is the Lyapunov
function of $\mathbf{x}'(t)=\mathbf{b}(\mathbf{x})$.

Our probabilistic derivation shows that the decomposition of the vector
field $\mathbf{b}$ in Eq. \eref{eq: b decomposition} is indeed
general. The orthogonality condition discussed in Ref. \cite{qian_zeroth_2014}
is not needed. An interesting direction for future research is to
explore the existence of an orthogonal decomposition $-\mathbf{D}\tilde{\nabla}\Phi-\mathbf{Q}\tilde{\nabla}\Phi$
or a divergence-free decomposition $-\mathbf{D}\tilde{\nabla}\Phi+\tilde{\nabla}\times\mathbf{Q}$
in curvilinear coordinates where $\tilde{\nabla}$ is the new operator
defined accordingly. Two theorems connecting Ao's perturbations with
FW's perturbations of deterministic dynamics via coordinate transformations
are presented in the supplementary materials.

We note that the generally-valid gradient form $\mathbf{x}'(t)=-\left(\mathbf{D}+\mathbf{Q}\right)\nabla\varphi$
for a system with a steady state does not preclude the deterministic
dynamics to have attractors with nonzero velocity, \emph{e.g.} limit
cycle or strange attractor. Our derivation shows that while $\nabla\varphi=0$
on the attractor, $\mathbf{Q}$ actually diverges on the attractor
such that $-\mathbf{Q}\nabla\varphi=\mathbf{x}'(t)=O(1)$. This is
shown by taking $\epsilon\rightarrow0$ of Eq. \eref{eq: SDE of random perturbation}
to arrive {at} $\boldsymbol{\gamma}=-\mathbf{Q}\nabla\varphi$ with $\mathbf{Q}<\infty$
and $\nabla\varphi>0$ in a deleted neighborhood of the attractor,
and then using the continuity of $\mathbf{b}+\mathbf{D}\nabla\varphi=\boldsymbol{\gamma}$
near the attractor to draw the conclusion. A concrete example can
be found in Ref. \cite{zhu_limit_2006}. Here we provide a proof with
the method of random perturbation based on the bivectorial formalism
we introduced.

Finally, the notion of cycle velocity presents a nice physical picture
of the conjugate process \cite{yang_unified_2020}. The conjugate
process, as a notion of time reversal that reverses the stationary
flux $\mathbf{J}^{*}$, corresponds to the transpose of the anti-symmetric
matrix $\mathbf{Q}$, which is equivalent to reversing all the cycles. 

\ack{The authors thank Hong Qian for his guidance and the many helpful
discussions he had with the authors. The authors also thank Pin Ao
and Erin Angelini for the many helpful discussions and feedback on
our manuscript. We also thank the two anonymous reviewers for their comments and suggestions.}

\appendix

\section{Geometrical Meaning of Simple Bivectors}
In 3-D, we have the notion of a signed area spanned by two vectors
$\mathbf{u}$ and $\mathbf{v}$ as a vector by using the right-hand
rule, conventionally denoted as $\mathbf{u}\times\mathbf{v}$. This
notion of a signed area can be generalized to general $n$-dimension
with the notion of a wedge product $\wedge$ in geometrical algebra.
The new object $\mathbf{u}\wedge\mathbf{v}$ is called a \emph{simple}
\emph{bivector. }In 3-D, a simple bivector can be represented by a
vector. For dimension higher than three, that is no longer possible.
The fundamental reason is when assigning a ``direction'' for a bivector
by using the right-hand rule in dimension higher than 3, there is
more than one dimension that is perpendicular to the plane spanned
by $\mathbf{u}\wedge\mathbf{v}.$ Even though we can not represent
it as a vector, we can still think of it as a planar object.

The linear combination of simple bivector is a bivector. In dimension smaller than three (including three), all bivectors are simple. 
That is, all bivector can be represented by the wedge product of two vectors. 
For dimension higher than 3, that is no longer true. 
There are bivectors that can {not} be represented by a simple wedge product of two vectors.
The geometric meaning we presented here is only for simple bivectors.

The wedge product from it's definition have several properties. For
example, for two vectors in $\mathbb{R}^{n}$, we have $\mathbf{u}\wedge\mathbf{v}=-\mathbf{v}\wedge\mathbf{u}$;
$\alpha\mathbf{u}\wedge\beta\mathbf{v}=\alpha\beta\mathbf{u}\wedge\mathbf{v}$
where $\alpha,\beta\in\mathbb{R}$ are scalars; $\mathbf{u}\wedge\mathbf{u}=\mathbf{0}$;
and $\left(\mathbf{u}+\alpha\mathbf{v}\right)\wedge\mathbf{v}=\mathbf{u}\wedge\mathbf{v}$.
Besides these rather straightforward one, we also have the nontrivial
distributive property
\[
\left(\mathbf{u}+\mathbf{v}\right)\wedge\mathbf{w}=\mathbf{u}\wedge\mathbf{w}+\mathbf{v}\wedge\mathbf{w}.
\]
This very property leads to a (anti-symmetric) matrix representation
of bivector and also a Pythogorean theorem for perpendicular $\mathbf{u},\mathbf{v},$
and $\mathbf{w}$.

First, let us show that area actually satisfies Pythogorean theorem.
Let us denote $\Vert\mathbf{u}\wedge\mathbf{v}\Vert$ as the area
indicated by the bivector $\mathbf{u}\wedge\mathbf{v}.$ Then, from
basic geometry, we have 
\[
\Vert\mathbf{u}\wedge\mathbf{v}\Vert^{2}=\Vert\mathbf{u}\Vert^{2}\Vert\mathbf{v}\Vert^{2}-\left(\mathbf{u}\cdot\mathbf{v}\right)^{2}.
\]
Now, applying this to $\Vert\left(\mathbf{u}+\mathbf{v}\right)\wedge\mathbf{w}\Vert^{2}$
for perpendicular vectors: $\mathbf{u}\bot\mathbf{v}$, $\mathbf{v}\bot\mathbf{w}$, and
$\mathbf{u}\bot\mathbf{w}$, we get 
\begin{eqnarray}
\Vert\left(\mathbf{u}+\mathbf{v}\right)\wedge\mathbf{w}\Vert^{2} 
&=\Vert\mathbf{u}+\mathbf{v}\Vert^{2}\Vert\mathbf{w}\Vert^{2}-\left[\left(\mathbf{u}+\mathbf{v}\right)\cdot\mathbf{w}\right]^{2}\\
&=\left(\Vert\mathbf{u}\Vert^{2}+\Vert\mathbf{v}\Vert^{2}\right)\Vert\mathbf{w}\Vert^{2}=\Vert\mathbf{u}\wedge\mathbf{w}\Vert^{2}+\Vert\mathbf{v}\wedge\mathbf{w}\Vert^{2}.
\end{eqnarray}
This can be used to derive, for example, the area of a triangle with
three apexes $\left(a,0,0\right),\left(0,b,0\right),$ and $\left(0,0,c\right)$.
By applying the Pythogoream theorem twice, one would get the triangular
area given by $\sqrt{\left(\frac{ab}{2}\right)^{2}+\left(\frac{bc}{2}\right)^{2}+\left(\frac{ca}{2}\right)^{2}}$.

With the distributive property, the simple bivector $\mathbf{u}\wedge\mathbf{v}$
spanned by the two general vectors $\mathbf{u}$ and $\mathbf{v}$
in $\mathbb{R}^{n}$ has a orthogonal decomposition into $n\left(n-1\right)/2$
orthogonal unit signed area $\mathbf{e}_{i}\wedge\mathbf{e}_{j}$
and corresponding components $\left(u_{i}v_{j}-u_{j}v_{i}\right)$,
\begin{eqnarray}
\mathbf{u}\wedge\mathbf{v} & =\left(\sum_{i=1}^{n}u_{i}\mathbf{e}_{i}\right)\wedge\left(\sum_{j=1}^{n}v_{j}\mathbf{e}_{j}\right)=\sum_{i<j}\left(u_{i}v_{j}-u_{j}v_{i}\right)\mathbf{e}_{i}\wedge\mathbf{e}_{j}.
\end{eqnarray}
This shows that $\mathbf{u}\wedge\mathbf{v}$ can be represented by
an \emph{anti-symmetric matrix} with components $\left(u_{i}v_{j}-u_{j}v_{i}\right).$ 

It can be shown that $\Vert\mathbf{u}\wedge\mathbf{v}\Vert^{2}=\left(\mathbf{u}\cdot\mathbf{u}\right)\left(\mathbf{v}\cdot\mathbf{v}\right)-\left(\mathbf{u}\cdot\mathbf{v}\right)^{2}=\sum_{i<j}\left(u_{i}v_{j}-u_{j}v_{i}\right)^{2}$.
This implies that the inner product between two bivectors $\mathbf{A}$
and $\mathbf{B}$, which can be represented by anti-symmetric matrices,
is just half of the Frobenius product of two matrices, $\mathbf{A}\cdot\mathbf{B}=\sum_{i<j}A_{ij}B_{ij}=\frac{1}{2}\sum_{i,j}A_{ij}B_{ij}$.
It is the sum of all the $n\left(n-1\right)/2$ multiplied components

The inner product can tell us whether two simple bivectors are ``perpendicular''
or not. Two perpendicular areas are only defined with a shared edge.
Geometrically, we would expect $\mathbf{u}\bot\mathbf{v}=0\Leftrightarrow\left(\mathbf{u}\wedge\mathbf{w}\right)\cdot\left(\mathbf{v}\wedge\mathbf{w}\right)=0$.
This can be seen by the computation below,
\begin{eqnarray}
&\left(\mathbf{u}\wedge\mathbf{w}\right)\cdot\left(\mathbf{v}\wedge\mathbf{w}\right) 
=\sum_{i<j}\left(u_{i}w_{j}-u_{j}w_{i}\right)\left(v_{i}w_{j}-v_{j}w_{i}\right)\\
 & =\frac{1}{2}\sum_{i,j}\left(u_{i}w_{j}-u_{j}w_{i}\right)\left(v_{i}w_{j}-v_{j}w_{i}\right)\\
 & =\frac{1}{2}\sum_{i,j}\left[u_{i}v_{i}w_{j}^{2}+u_{j}v_{j}w_{i}^{2}-\left(u_{i}v_{j}-u_{j}v_{i}\right)w_{i}w_{j}\right]\\
 & =\left(\sum_{i}u_{i}v_{i}\right)\sum_{j}w_{j}^{2}.
\end{eqnarray}

\section{Ao's perturbation of FPE and HJE}
In this {appendix}, we focus on { some features and novelty about} Ao's perturbations of Fokker-Planck equations (FPEs) and { its} corresponding Hamilton-Jacobi equations (HJEs). In general, Fokker-Planck equations with small noises $\epsilon \mD(\vx)$ is considered as 
\numparts
\begin{eqnarray}
\label{eqFPE}
  \frac{\partial p_\epsilon(\vx,t)}{\partial t}
   &=& -\nabla\cdot\vJ_\epsilon(\vx,t), 
\\
    \vJ_\epsilon(\vx,t) &=& \vb_\epsilon(\vx) p_\epsilon(\vx,t) - \epsilon\mD(\vx)\nabla p_\epsilon(\vx,t),
\end{eqnarray}
\endnumparts
in which $\vb_\epsilon(\vx)$ has a general decomposition 
\begin{equation}
\mathbf{b}_{\epsilon}(\vx)=-\epsilon\mathbf{D}(\vx)\nabla\Phi_{\epsilon}(\vx)-\mathbf{Q}_{\epsilon}(\vx)\nabla\Phi_{\epsilon}(\vx)+\nabla\times\mathbf{Q}_{\epsilon}(\vx).\label{eq: b_epsilon}
\end{equation}
{ Based on Eq. \eref{eq: b_epsilon}, as we mentioned in Sect. 4.3, Yin and Ao suggested a particular scaling $\mathbf{Q}_\epsilon = \epsilon \mathbf{Q}$ and $\Phi_{\epsilon} = \phi / \epsilon$ which leads to
\begin{equation} \label{Ao drift}
\vb_\epsilon(\vx) = - \mD(\vx) \nabla \phi(\vx) - \mQ(\vx) \nabla \phi(\vx) + \epsilon \nabla\times\mathbf{Q}(\vx).\label{eq: b_epsilon2}
\end{equation}}

{To find the asymptotic series solutions of Eq. \eref{eqFPE}, we apply the WKB ansatz with another $\epsilon$-dependent function $\varphi_\epsilon(\vx, t)$} 
\begin{eqnarray} \label{wkb_app}
    p_{\epsilon}(\vx, t) =  \exp \left[ - \varphi_\epsilon(\vx, t)/ \epsilon \right], \\ \varphi_\epsilon(\vx, t) = \varphi(\vx, t) - a(t) \epsilon\ln \epsilon - \epsilon\ln\omega(\vx, t)  + o(\epsilon), 
\end{eqnarray}
in which the leading order term $\varphi(\vx, t)$ is called a time-dependent rate function of large deviation, $a(t) \epsilon\ln \epsilon$ is from the normalization factor, and $ \ln \omega(\vx, t)$ is the next order term.  By plugging this WKB ansatz into Eq. \eref{eqFPE}, we can show that $\varphi(\vx, t)$  satisfies the HJE
\numparts
\label{eqHJE}
\begin{eqnarray}
  \frac{\partial \varphi(\vx,t)}{\partial t}
   &=& -\bf \gamma(\vx, t) \cdot \nabla\varphi(\vx,t) , 
\\
    \bf \gamma(\vx, t) &=& \mD(\vx)\nabla \varphi(\vx,t) + \vb(\vx),
\end{eqnarray}
\endnumparts
{ in which $ \vb(\vx) = \lim_{\epsilon \rightarrow 0} \vb_\epsilon(x)$.} Note that this HJE can have non-smooth solution after certain finite time $T > 0$ by studying the characteristics of it. In mathematics, the long-term behavior of $\varphi(\vx,t)$ can be understood by {\em viscosity solutions} with the method of vanishing viscosity: it turns the original nonlinear first-order PDE into a quasilinear parabolic PDE by introducing a small term involving {$\epsilon \nabla  \nabla \varphi(\vx,t)$ \cite{evans_partial_2010}.} This mathematical technique would tell us why there is always a globally smooth potential $\phi$ in Ao's perturbation as follows. 

Let us apply the “pre-expansion” form of WKB ansatz $e^{-\varphi_\epsilon(\vx, t) /\epsilon }$ in Eq. \eref{wkb_app}  directly to the Fokker-Planck equations with Eq. \eref{Ao drift}. We then obtain that 
\numparts
\begin{eqnarray} \label{HJE-AO-2}
\frac{\partial\varphi_\epsilon(\vx, t) }{\partial t} &= - \mD\nabla\varphi_\epsilon \cdot \nabla\varphi_\epsilon + \mD\nabla\phi \cdot \nabla\varphi_\epsilon + \mQ\nabla\phi \cdot \nabla\varphi_\epsilon \\ \label{HJE-AO-3}
&+ \epsilon\big[  (\nabla\times \mQ)\cdot \left( \nabla \varphi_\epsilon -  \nabla \phi\right) + \nabla \cdot \left(\mD \nabla\varphi_\epsilon - \mD \nabla\phi\right)  \big],
\end{eqnarray}
\endnumparts
which provides a natural {\em viscous Hamilton–Jacobi equation} with the diffusion term involving $\epsilon \nabla  \nabla \varphi(\vx,t)$ in Eq. \eref{HJE-AO-3}. {And we can check that $ \phi $ is an invariant solution for all  $\epsilon > 0$, i.e, 
\begin{eqnarray} \label{eq: stat-soln-Ao-a}
\phi(\vx) = \lim_{t\rightarrow \infty}\varphi_\epsilon(\vx, t) \quad \quad \forall \epsilon > 0.
\end{eqnarray}
Therefore, we can take limit of $\epsilon$ goes to zero on the both sides of Eq. \eref{eq: stat-soln-Ao-a} to get
\begin{eqnarray} \label{eq: stat-soln-Ao-b}
   \phi(\vx) = \lim_{\epsilon \rightarrow 0} \lim_{t\rightarrow \infty}\varphi_\epsilon(\vx, t) = - \lim_{\epsilon \rightarrow 0} \lim_{t\rightarrow \infty} \epsilon \ln p_\epsilon(\vx,t).
\end{eqnarray}
Note that this vanishing viscosity is independent of $\epsilon$, so it avoids losing control over the smoothness of the function while taking limit of $\epsilon \rightarrow 0$. Furthermore, by this order of taking limits ($\lim_{\epsilon \rightarrow 0} \lim_{t\rightarrow \infty}$) of $\ln p_\epsilon(\vx,t)$ in Eq. \eref{eq: stat-soln-Ao-b}, $\phi$ portrays the landscape for the whole space $\mathbb{R}^n$ \cite{zhou_construction_2016}. Therefore, Ao's perturbation of FPE arises a {\em global} and {\em smooth} potential $\phi$.}

For the invariant density, again,
plugging the stationary WKB ansatz 
\begin{eqnarray} \label{wkb-app}
 \pi_{\epsilon}(\vx) = \exp \left[ - \varphi_\epsilon(\vx) / \epsilon \right], \\  \varphi_\epsilon(\vx) = \varphi(\vx) - a \epsilon \ln \epsilon - \epsilon  \ln \omega(\vx) + o(\epsilon), 
\end{eqnarray}
into the stationary Eq. \eref{eqFPE}, we  can show that $\varphi(\vx)$ and $\omega(\vx)$ satisfies a system of three equations in QCY's perturbation \cite{qian_kinematic_2020}
\numparts
\label{hq-eq}
\begin{eqnarray}
  & \vb(\vx) = -\mD(\vx)\nabla\varphi(\vx) + \vgamma(\vx),
\\[3pt] \label{ortho-1}
   &  \nabla\varphi(\vx)\cdot \vgamma(\vx) = 0,
\\[3pt]
  &\nabla\cdot\big( \omega(\vx) \vgamma(\vx) \big) 
    = -\nabla\omega(\vx) \cdot \mD(\vx)\nabla\varphi(\vx). \label{omega-1}
\end{eqnarray}
\endnumparts
On the other hand, { by plugging the WKB ansatz \eref{wkb-app} and Eq. \eref{eq: stat-soln-Ao-a} into stationary Eq.\eref{eqFPE} with Eq. \eref{Ao drift},} we  obtain the other system of three equations
\numparts
\label{hq-eq-2}
\begin{eqnarray}
  & \vb(\vx) = -\mD(\vx)\nabla\phi(\vx) + \vgamma(\vx),
\\[3pt] \label{ortho-2}
   &  \nabla\phi(\vx)\cdot\vgamma(\vx) = 0,
\\[3pt]
  &\nabla \omega(\vx) \cdot \vgamma(\vx)
    = -\nabla\omega(\vx) \cdot \mD(\vx)\nabla\phi(\vx). \label{omega-2}
\end{eqnarray}
\endnumparts

Those two sets of equations provide a similar but different geometric interpretations of the vector field $\vgamma$. The motions of $\vgamma$ are both restricted on the level set of $\varphi$ or $\phi$ by the orthogonality \eref{ortho-1} or \eref{ortho-2}, respectively. {As we discussed in Sect. 4.3, by the unique orthogonal decomposition theorem by Freidlin and Wentzell, $\varphi$ and $\phi$ have the same shape in a certain domain.} On the other hand, we can find that the main difference is in the equation of the prefactor $\omega$, which represents local measure for the phase space volume, so it is also called the phase space factor \cite{kampen_stochastic_2007}. In QCY's work, if we choose $\omega(\vx) = 1$, i.e., {the prefactor is uniform in the phase space}, then the vector field $\vgamma(\vx)$ has to be volume preserving by Eq. \eref{omega-1}. {However, in Ao' perturbations, the particular choice of $\omega(\vx) = 1$ is not able to give us a volume preserving $\vgamma$ due to the following reason: Once we choose $\omega(\vx) = 1$ and plug it into Eq. \eref{omega-2}, since it is a constant,  Eq. \eref{omega-2} is always $0 =0$ with an undetermined vector field $\vgamma$.} In other words, in Ao's perturbation, the current term $\vgamma$ is allowed to have a conserved $\phi$ in the phase space with a uniform prefactor, but $\vgamma$ is still not volume-preserving; this is very special in contradistinction to the common perturbation of small-noises Fokker-Planck equations.  

\section{Change of coordinate}
Given a stochastic process 
\begin{equation}
\mathrm{d}\mathbf{X}_{t}=\left(\mathbf{b}(\mathbf{X}_{t})+\nabla\cdot\mathbf{D}(\mathbf{X}_{t})\right)\mathrm{d}t+\sqrt{2\mathbf{D}(\mathbf{X}_{t})}\mathrm{d}\mathbf{W}_{t}, \label{eq: SDE-A}
\end{equation}
we have shown that the {vector field} $\vb$ has a general decomposition 
\begin{eqnarray}
    \vb = - \mD \nabla \Phi - \mQ \nabla \Phi  + \nabla \times \mQ.
\end{eqnarray}
{ For a random perturbation of deterministic dynamics such as
\begin{eqnarray}
  \mathrm{d}\mathbf{X}_{t}= \left(\mathbf{b}_{\epsilon}(\mathbf{X}_{t})+\epsilon \nabla\cdot\mathbf{D}(\mathbf{X}_{t})\right)\mathrm{d}t+\sqrt{2\epsilon\mathbf{D}(\mathbf{X}_{t})}\mathrm{d}\mathbf{W}_{t}, 
\end{eqnarray}
suggested by Yin and Ao \cite{yin_existence_2006}, the {vector field} $\vb_\epsilon$ can be particularly decomposed as 
\begin{eqnarray}
    \vb_\epsilon = - \mD \nabla \phi - \mQ \nabla \phi  + \epsilon \nabla \times \mQ,
\end{eqnarray}
and this decomposition has been shown having several nice properties. But some questions arise: What is the relation between this type of perturbation and the perturbation by Freidlin-Wentzell theory? Does there exist transformations between the two types of perturbations?}

To answer the above questions, we have to rigorously give definitions of those two types of perturbations:

\paragraph{\textup{\textbf{Definition} (Ao-type perturbation of  processes):}}
A random process $\vX_t$ satisfies the SDE
\begin{eqnarray} \label{random-Ao}
     \rd \vX_t =& \big[- \big(\mD(\vX_t) + \mQ(\vX_t) \big)\nabla\phi(\vX_t)   +  \epsilon \nabla \cdot \mD(\vX_t) \nonumber
     \\ &+\epsilon \nabla \times  \mQ(\vX_t)\big) \big] \rd t + \sqrt{2\epsilon\mD(\vX_t)} \rd \mathbf{W}_t, 
\end{eqnarray}
in which the matrix $\mD$ is symmetric  and the matrix $\mQ$ is anti-symmetric, then $\vX_t$ is called Ao-type perturbation of processes.

\paragraph{\textup{\textbf{Definition} (FW-type perturbation of  processes):}}
A random process $\vY_t$ satisfies the SDE
\begin{eqnarray} \label{random-gradient}
    \rd \mathbf{Y}_t= -\bf k(\vY_t) \rd t + \sqrt{2\epsilon \mathbf{S}(\mathbf{Y}_t)}\rd \mathbf{W}_t.
\end{eqnarray}
in which $\bf S$ is a symmetric matrix, then $\vY_t$ is called FW-type perturbation of processes.

Let us consider the following deterministic dynamics 
\begin{eqnarray} \label{fw-ode-1}
    \rd \vy(t) = - \bf k (\vy) \rd t,
\end{eqnarray}
in which the vector field $\bf k$ might {\em not} be a gradient flow. If we can find a bijective  map, e.g., a coordinate transformation, such that the vector field $\bf k$ becomes a gradient flow in the Euclidean norm, then the following two theorems show that there exists a one-to-one correspondence  between a subset of FW-type  perturbations and  a subset of Ao-type perturbations of the dynamics \eref{fw-ode-1}. Theorem \hyperref[FW-to-Ao]{1} gives a transformation of FW-type to Ao-type processes; and Theorem \hyperref[Ao-to-FW]{2} gives a transformation of Ao-type to FW-type processes: 

\paragraph{\textbf{\textup{Theorem 1} } } 
\label{FW-to-Ao}
Let $\vY_t$ be a FW-type perturbation process. Assume there exists a bijective function $\vf \in C^2(\mathbb{R}^n)$ such that {the composition of functions
$\left( \bf k \circ \vf^{-1} \right)$}is a gradient flow in the Euclidean norm and 
\begin{eqnarray}
    \mS(\vy) = \frac{1}{2}\big[ \mA(\vy)^{-1} + \mA(\vy)^{-T}\big], 
\end{eqnarray}
where $\mA$ is the Jacobian matrix of $\vf$.
Then $\vX_t = \vf(\vY_t)$ has to be a Ao-type perturbation of process with
\begin{eqnarray}
  \nabla \phi(\vx) &=&  \left( \bf k \circ \vf^{-1} \right)(\vx), \\ 
  \mD(\vx) &=& \frac{1}{2}\big[ \mA\left(\vf^{-1}(\vx)\right) + \mA^T\left(\vf^{-1}(\vx)\right) \big], \\
  \mQ(\vx) &=& \frac{1}{2}\big[ \mA\left(\vf^{-1}(\vx)\right) - \mA^T\left(\vf^{-1}(\vx)\right) \big].
\end{eqnarray}
Furthermore, the process $\vX_t$ has the invariant density 
$e^{- \frac{\phi(\vx)}{\epsilon}}.$

\paragraph{\textbf{\textup{Theorem 2} } } \label{Ao-to-FW}
Let $\vX_t$ be a Ao-type perturbation process. Assume there exists a bijective function $\vf \in C^2(\mathbb{R}^n)$  such that 
\begin{eqnarray}
   \mD(\vx) + \mQ(\vx) = \mA( \vf^{-1}(\vx)),
\end{eqnarray}
where $\mA$ is the Jacobian matrix of $\vf$.
Then $\vY_t = \vf^{-1} (\vX_t)$ has to be a FW-type perturbation process with
\begin{eqnarray}
\bf k (\vy) &=&  \left( \nabla \phi  \circ \vf \right)(\vy), \label{map-k-varphi} \\
  \mS(\vy) &=&  \frac{1}{2}\big[ \mA(\vy)^{-1} + \mA(\vy)^{-T}\big]. 
\end{eqnarray}
Furthermore, the process $\vY_t$ has an invariant density 
$\omega(\vy)e^{- \frac{\varphi(\vf(\vy))}{\epsilon}},$
the prefactor $\omega(\vy)$ is the absolute value of the determinant of $\mA(\vy)$.

\paragraph{Proof}
The following proof covers both Theorem \hyperref[FW-to-Ao]{1} and \hyperref[Ao-to-FW]{2}. Let $\vf = (f_1, \cdots, f_n)$, $\vf$ is bijective, and $\vx = \vf(\vy)$, and $x_k = f_k(\vy), \ k=1, \cdots, n$. By multi-dimension It\^o formula, we have 
\begin{eqnarray} \label{ito-formula}
 \rd \vX_k(t) = \rd f_k(\vY_k(t))= \frac{\partial f_k(\vY)}{\partial y_i} \rd \vY_i    + \frac{1}{2} \frac{\partial^2 f_k(\vY)}{\partial y_i \partial y_j}\rd\vY_i\rd \vY_j. 
\end{eqnarray}
Plug \eref{random-gradient} into \eref{ito-formula}, we get
\begin{eqnarray}
    \rd \vX_k(t) &= - \frac{\partial f_k(\vY)}{\partial y_i} \bf k(\vY)_i \rd t + \frac{\partial f_k(\vY)}{\partial y_i} ( \sqrt{2\epsilon \bf S(\vY)} \rd \mathbf{B}_t )_i\nonumber \\
    &+ \frac{1}{2}\frac{\partial^2 f_k(\vY)}{\partial y_i \partial y_j}( \sqrt{2\epsilon \bf S(\vY)} \rd \mathbf{B}_t )_i( \sqrt{2\epsilon \bf S(\vY)} \rd \mathbf{B}_t )_j. 
\end{eqnarray}
To rewrite it in a vector-matrix form with, we have
\begin{eqnarray}
    \rd \vX(t) = &-\mA(\vY) \bf k (\vY)\rd t + \sqrt{2\epsilon}\big[ \mA(\vY)\sqrt{\bf S(\vY)}\big] \rd \mathbf{B}_t \nonumber \\
    &+ \epsilon \frac{1}{2} \mathbf{H} \bf S(\vY) \rd t, 
\end{eqnarray}
where $\mathbf{H}$ is a 3-rank tensor,  $(\mathbf{H})_{kji} = \frac{\partial^2 f_k(\vY)}{\partial y_i \partial y_j}$ and $\mathbf{H} \mathbf{S}$ follows tensor multiplication. By comparison term by term, we have a system of equations
\numparts
\begin{eqnarray} 
   \bf k (\vy) &=& \nabla \varphi(\vx)  \label{sys-0} \\
   \mA(\vy) &=& \mD(\vx) + \mQ(\vx), \label{sys-1} \\
   \mA(\vy)\sqrt{\bf S(\vy)} &=& \sqrt{\mD(\vx)}, \label{sys-2} \\
   \mathbf{H} \bf S(\vy) &=& \nabla \cdot \mD(\vx) + \nabla \times \mQ(\vx) \label{sys-3}.
\end{eqnarray}
\endnumparts
Therefore, if a map satisfies Eq. \eref{sys-0} - \eref{sys-3}, it will be a transformation between $\vY_t$ and $\vX_t$. 

Now, we are ready to prove Theorem \hyperref[FW-to-Ao]{1}: By the assumption $\left( \bf k \circ \vf^{-1} \right)$ is a gradient flow in the Euclidean norm, we can define  $\nabla \phi(\vx) =  \left( \bf k \circ \vf^{-1} \right)(\vx)$, which implies Eq. \eref{sys-0}.   By another assumption, \begin{eqnarray}
    \mS(\vy) = \frac{1}{2}\big[ \mA(\vy)^{-1} + \mA(\vy)^{-T}\big],
\end{eqnarray}
and let 
\numparts
\begin{eqnarray}
  \mD(\vx) &=& \frac{1}{2}\big[ \mA(\vy) + \mA^T(\vy)\big]= \frac{1}{2}\big[ \mA\left(\vf^{-1}(\vx)\right) + \mA^T\left(\vf^{-1}(\vx)\right) \big], \\
  \mQ(\vx) &=& \frac{1}{2}\big[ \mA(\vy) - \mA^T(\vy)\big]= \frac{1}{2}\big[ \mA\left(\vf^{-1}(\vx)\right) - \mA^T\left(\vf^{-1}(\vx)\right) \big],
\end{eqnarray}
\endnumparts
which is satisfied by the conditions of $\mD(\vx)$ and $\mQ(\vx)$, then we can check that the Eqs. \eref{sys-1} and \eref{sys-2} are satisfied.

In addition, the LHS of Eq. \eref{sys-3} can be written as 
\begin{eqnarray}
 \frac{1}{2} \sum_{i,j} \frac{\partial^2 f_k(\vy)}{\partial y_i \partial y_j} \mathbf{S}_{ij} &=  \frac{1}{2} \sum_{i,j} \frac{\partial (\mA(\vy))_{ki}}{\partial y_j}\big[(\mA(\vy)^{-1})_{ij} + (\mA(\vy)^{-1})_{ji} \big]\nonumber \\
 & = \frac{1}{2}  \left[ \sum_{j} \frac{\partial f_k(\vy)}{\partial y_j \partial x_j} + \sum_{i} \frac{\partial f_k(\vy)}{\partial y_i \partial x_i} \right] \nonumber \\
 &= \sum_{i} \frac{\partial f_k(\vy)}{\partial y_i \partial x_i},
\end{eqnarray}
and the RHS of Eq. \eref{sys-3} can be written as 
\begin{eqnarray}
 \sum_{i} \frac{\partial \left(\mD(\vx)_{ik} - \mQ(\vx)_{ik}\right)}{\partial x_i} &=  \sum_{i} \frac{\partial (\mA(\vy))_{ki}}{\partial x_i} = \sum_{i} \frac{\partial f_k(\vy)}{\partial y_i \partial x_i},
\end{eqnarray}
so we have LHS = RHS of \eref{sys-3}. Therefore,  Eq. \eref{sys-0} - \eref{sys-3} are satisfied, so $\vf$ maps $\vY_t$ to $\vX_t$,  which proves Theorem \hyperref[FW-to-Ao]{1}.

For Theorem \hyperref[Ao-to-FW]{2}, by $\vy=\vf^{-1}(\vx)$ and $\bf k (\vy) =  \left( \nabla \phi  \circ \vf \right)(\vy)$, we can get  Eq. \eref{sys-0}. Furthermore, by the assumption 
\begin{eqnarray} \label{d+q}
    \mD(\vx) + \mQ(\vx) = \mA( \vf^{-1}(\vx)),
\end{eqnarray}
and let 
\begin{eqnarray} \label{s=invd}
   \mS(\vy)=  \frac{1}{2}\big[ \mA(\vy)^{-1} + \mA(\vy)^{-T}\big],
\end{eqnarray}
we will obtain Eq. \eref{sys-1} and Eq. \eref{sys-2}. In addition, by \eref{d+q} and \eref{s=invd}, we can check that Eq. \eref{sys-3} is always hold as we proved for Theorem \hyperref[Ao-to-FW]{2}.

For the invariant density,  Ao-type perturbation of process $\vX_t$ has the invariant density 
$
     e^{- \frac{\phi(\vx)}{\epsilon}},
$ has been proved previously; FW-type invariant densitiy can be proved by the transformation of density function for $\vX_{\infty} = \vf(\vY_{\infty})$
\begin{eqnarray}
    \tilde{\pi}(\vy) \bigg | \frac{\rd \vx}{\rd \vy} \bigg | =   \pi(\vx), \quad \rm{where} \  \bigg | \frac{\rd \vx}{\rd \vy} \bigg | = \rm{det}(\mA).  
\end{eqnarray}
\null\hfill$\square$

\paragraph{Interpretations of Theorem \hyperref[FW-to-Ao]{1} and \hyperref[Ao-to-FW]{2}.}
{We start with a deterministic dynamics \eref{fw-ode-1} following the vector field $\bf k$ which might not be a gradient flow. Through a coordinate transformation, $\bf k \circ \vf^{-1}$ becomes a gradient flow. Then we consider a FW-type perturbation of the dynamics with a diffusion term related to the coordinate transformation as follows
\numparts
\begin{eqnarray}
    \rd \mathbf{Y}(t) &= -\bf k(\vY) \rd t + \sqrt{2\epsilon \mathbf{S}(\mathbf{Y})}\rd \mB_t , \\ \mS(\vy) &= \frac{1}{2}\big[ \mA(\vy)^{-1} + \mA(\vy)^{-T}\big].
\end{eqnarray}
\endnumparts
By Theorem \hyperref[FW-to-Ao]{1}, it shows that the random process $\vX_t = \vf(\vY_t)$ has to be a Ao-type process. 

Conversely, given any Ao-type process, we first assume that there exists a bijetive function $\vf$ such that
\begin{eqnarray}
   \mD(\vx) + \mQ(\vx) = \mA( \vf^{-1}(\vx)).
\end{eqnarray}
By Theorem \hyperref[Ao-to-FW]{2}, using the function $\vf^{-1}$ as a coordinate transformation, $\vY_t = \vf^{-1}(\vX_t)$, we can map the Ao-type process back to a FW-type process, i.e. the $O(\epsilon)$ term in the vector field can be killed by this transformation. Therefore, those two theorems provide a one-to-one map between a subset of Ao-type processes and a subset of FW-type processes. }

For the invariant density, Ao-type processes has the form $e^{- \frac{\phi(\vx)}{\epsilon}}$; the corresponding FW-type processes has the invariant density $\omega(\vy)e^{- \frac{\phi(\vf(\vy))}{\epsilon}},$
the prefactor $\omega(\vy)$ can be regarded as a weight due to the Jacobian of the transformation.

\section*{References}

\bibliographystyle{iopart-num}
\bibliography{Decomposition-rev2}

\providecommand{\newblock}{}
\begin{thebibliography}{10}
\expandafter\ifx\csname url\endcsname\relax
  \def\url#1{{\tt #1}}\fi
\expandafter\ifx\csname urlprefix\endcsname\relax\def\urlprefix{URL }\fi
\providecommand{\eprint}[2][]{\url{#2}}

\bibitem{berlinski_infinite_2008}
Berlinski D 2008 {\em Infinite {Ascent}: {A} {Short} {History} of
  {Mathematics}\/} reprint edition ed (New York, NY: Modern Library) ISBN
  978-0-8129-7871-1

\bibitem{kubo_fluctuation_1973}
Kubo R, Matsuo K and Kitahara K 1973 {\em J Stat Phys\/} {\bf 9} 51--96 ISSN
  1572-9613 \urlprefix\url{https://doi.org/10.1007/BF01016797}

\bibitem{graham_non-equilibrium_1983}
Graham R and Schenzle A 1983 {\em Z. Physik B - Condensed Matter\/} {\bf 52}
  61--68 ISSN 1431-584X \urlprefix\url{https://doi.org/10.1007/BF01305899}

\bibitem{nicolis_comment_1977}
Nicolis G and Lefever R 1977 {\em Physics Letters A\/} {\bf 62} 469--471 ISSN
  0375-9601
  \urlprefix\url{http://www.sciencedirect.com/science/article/pii/037596017790069X}

\bibitem{seifert_stochastic_2012}
Seifert U 2012 {\em Rep. Prog. Phys.\/} {\bf 75} 126001 ISSN 0034-4885
  \urlprefix\url{http://stacks.iop.org/0034-4885/75/i=12/a=126001}

\bibitem{thompson_nonlinear_2016}
Thompson L~F and Qian H 2016 {\em Entropy\/} {\bf 18} 309 ISSN 1099-4300
  \urlprefix\url{http://arxiv.org/abs/1605.08071}

\bibitem{yang_unified_2020}
Yang Y~J and Qian H 2020 {\em Phys. Rev. E\/} {\bf 101} 022129
  \urlprefix\url{https://link.aps.org/doi/10.1103/PhysRevE.101.022129}

\bibitem{wang_potential_2008}
Wang J, Xu L and Wang E 2008 {\em PNAS\/} {\bf 105} 12271--12276 ISSN
  0027-8424, 1091-6490
  \urlprefix\url{https://www.pnas.org/content/105/34/12271}

\bibitem{ge_extended_2009}
Ge H 2009 {\em Phys. Rev. E\/} {\bf 80} 021137
  \urlprefix\url{https://link.aps.org/doi/10.1103/PhysRevE.80.021137}

\bibitem{esposito_three_2010}
Esposito M and Van~den Broeck C 2010 {\em Phys. Rev. Lett.\/} {\bf 104} ISSN
  0031-9007, 1079-7114
  \urlprefix\url{https://link.aps.org/doi/10.1103/PhysRevLett.104.090601}

\bibitem{yang_bivectorial_2021}
Yang Y~J and Qian H 2021 {\em J Stat Phys\/} {\bf 182} 46 ISSN 1572-9613
  \urlprefix\url{https://doi.org/10.1007/s10955-021-02723-3}

\bibitem{ao_potential_2004}
Ao P 2004 {\em J. Phys. A: Math. Gen.\/} {\bf 37} L25--L30 ISSN 0305-4470
  \urlprefix\url{https://doi.org/10.1088%2F0305-4470%2F37%2F3%2Fl01}

\bibitem{yin_existence_2006}
Yin L and Ao P 2006 {\em J. Phys. A: Math. Gen.\/} {\bf 39} 8593--8601 ISSN
  0305-4470 \urlprefix\url{https://doi.org/10.1088%2F0305-4470%2F39%2F27%2F003}

\bibitem{freidlin_random_2012}
Freidlin M~I and Wentzell A~D 2012 {\em Random {Perturbations} of {Dynamical}
  {Systems}\/} 3rd ed Grundlehren der mathematischen {Wissenschaften} (Berlin
  Heidelberg: Springer-Verlag) ISBN 978-3-642-25846-6
  \urlprefix\url{https://www.springer.com/gp/book/9783642258466}

\bibitem{qian_kinematic_2020}
Qian H, Cheng Y~C and Yang Y~J 2020 {\em EPL\/} {\bf 131} 50002 ISSN 0295-5075
  \urlprefix\url{https://doi.org/10.1209%2F0295-5075%2F131%2F50002}

\bibitem{hong_representations_2020}
Hong L, Qian H and Thompson L~F 2020 {\em J Comput Appl Math\/}  112842 ISSN
  0377-0427
  \urlprefix\url{http://www.sciencedirect.com/science/article/pii/S0377042720301333}

\bibitem{feller_general_1954}
Feller W 1954 {\em ANN MATH\/} {\bf 60} 417--436 ISSN 0003-486X
  \urlprefix\url{https://www.jstor.org/stable/1969842}

\bibitem{jiang_mathematical_2004}
Jiang D~Q, Qian M and Qian M~P 2004 {\em Mathematical {Theory} of
  {Nonequilibrium} {Steady} {States}: {On} the {Frontier} of {Probability} and
  {Dynamical} {Systems}\/} 2004th ed (Berlin ; New York: Springer) ISBN
  978-3-540-20611-8

\bibitem{lebowitz_gallavotti-cohen-type_1999}
Lebowitz J~L and Spohn H 1999 {\em J Stat Phys\/} {\bf 95} 333--365 ISSN
  1572-9613 \urlprefix\url{https://doi.org/10.1023/A:1004589714161}

\bibitem{qian_relative_2001}
Qian H 2001 {\em Phys. Rev. E\/} {\bf 63} 042103
  \urlprefix\url{https://link.aps.org/doi/10.1103/PhysRevE.63.042103}

\bibitem{wang_mathematical_2020}
Wang Y and Qian H 2020 {\em J Stat Phys\/} {\bf 179} 808--837 ISSN 1572-9613
  \urlprefix\url{https://doi.org/10.1007/s10955-020-02556-6}

\bibitem{qian_decomposition_2013}
Qian H 2013 {\em J. Math. Phys.\/} {\bf 54} 053302 ISSN 0022-2488
  \urlprefix\url{https://aip.scitation.org/doi/abs/10.1063/1.4803847}

\bibitem{qian_zeroth_2014}
Qian H 2014 {\em Phys. Lett. A\/} {\bf 378} 609--616 ISSN 0375-9601
  \urlprefix\url{http://www.sciencedirect.com/science/article/pii/S0375960113011766}

\bibitem{qian_thermodynamics_2015}
Qian H 2015 {\em Eur. Phys. J. Spec. Top.\/} {\bf 224} 781--799 ISSN 1951-6401
  \urlprefix\url{https://doi.org/10.1140/epjst/e2015-02427-6}

\bibitem{graham_covariant_1977}
Graham R 1977 {\em Z Physik B\/} {\bf 26} 397--405 ISSN 1431-584X
  \urlprefix\url{https://doi.org/10.1007/BF01570750}

\bibitem{ding_covariant_2020}
Ding M, Tu Z and Xing X 2020 {\em Phys. Rev. Research\/} {\bf 2} 033381
  \urlprefix\url{https://link.aps.org/doi/10.1103/PhysRevResearch.2.033381}

\bibitem{shi_relation_2012}
Shi J, Chen T, Yuan R, Yuan B and Ao P 2012 {\em J Stat Phys\/} {\bf 148}
  579--590 ISSN 1572-9613
  \urlprefix\url{https://doi.org/10.1007/s10955-012-0532-8}

\bibitem{huang_processes_2017}
Huang S, Li F, Zhou J~X and Qian H 2017 {\em J R Soc Interface\/} {\bf 14} ISSN
  1742-5662

\bibitem{zhu_limit_2006}
Zhu X~M, Yin L and Ao P 2006 {\em Int. J. Mod. Phys. B\/} {\bf 20} 817--827
  ISSN 0217-9792
  \urlprefix\url{https://www.worldscientific.com/doi/10.1142/S0217979206033607}

\bibitem{zhou_construction_2016}
Zhou P and Li T 2016 {\em J. Chem. Phys.\/} {\bf 144} 094109 ISSN 0021-9606
  \urlprefix\url{https://aip.scitation.org/doi/10.1063/1.4943096}

\bibitem{nolting_balls_2015}
Nolting B~C and Abbott K~C 2015 {\em Ecology\/}  15--1047.1 ISSN 0012-9658
  \urlprefix\url{http://doi.wiley.com/10.1890/15-1047.1}

\bibitem{tang_potential_2017}
Tang Y, Yuan R, Wang G, Zhu X and Ao P 2017 {\em Sci Rep\/} {\bf 7} 15762 ISSN
  2045-2322 \urlprefix\url{http://www.nature.com/articles/s41598-017-15889-2}

\bibitem{cheng_stochastic_2021}
Cheng Y~C and Qian H 2021 {\em J Stat Phys\/} {\bf 182} 47 ISSN 1572-9613
  \urlprefix\url{https://doi.org/10.1007/s10955-021-02724-2}

\bibitem{kwon_structure_2005}
Kwon C, Ao P and Thouless D~J 2005 {\em PNAS\/} {\bf 102} 13029--13033 ISSN
  0027-8424, 1091-6490
  \urlprefix\url{https://www.pnas.org/content/102/37/13029}

\bibitem{evans_partial_2010}
Evans L~C 2010 {\em Partial {Differential} {Equations}: {Second} {Edition}\/}
  2nd ed (Providence, R.I: American Mathematical Society) ISBN
  978-0-8218-4974-3

\bibitem{kampen_stochastic_2007}
Kampen N~G~V 2007 {\em Stochastic {Processes} in {Physics} and {Chemistry}\/}
  3rd ed (Amsterdam ; Boston: North Holland) ISBN 978-0-444-52965-7

\end{thebibliography}

\end{document}